# Towards coherent spin precession in pure-spin current

Hiroshi Idzuchi[1,2], Yasuhiro Fukuma[2,3*], and YoshiChika Otani[1,2]

[1]*Institute for Solid State Physics, University of Tokyo, Kashiwa 277-8581, Japan*
[2]*Advanced Science Institute, RIKEN, 2-1, Wako 351-0198, Japan*
[3] *Frontier Research Academy for Young Researchers, Kyushu Institute of Technology, Iizuka 820-8502, Japan*

Non-local spin injection in lateral spin valves generates a pure spin current which is a diffusive flow of spins (i.e. spin angular momentums) with no net charge flow. The diffusive spins lose phase coherency in precession while undergoing frequent collisions and these events lead to a broad distribution of the dwell time in a transport channel between the injector and the detector. Here we show the lateral spin-valves with dual injectors enable us to detect a genuine in-plane precession signal from the Hanle effect, demonstrating the phase coherency in the in-plane precession is improved with an increase of the channel length. The coherency in the spin precession shows a universal behavior as a function of the normalized separation between the injector and the detector in material-independent fashion for metals and semiconductors including graphene.

*yfukuma@riken.jp

(**Introduction**)

Diffusion is a transport process driven by gradients in the concentration of particles in random motion with undergoing collisions such as molecules in water, disorder in crystals, and electrons and holes in a semiconductor[1]. Recent advance in nano-scale fabrication technology has opened up a possibility for studying such a diffusive transport of accumulated spins in a nonmagnet by means of nonlocal spin injection[2-5]. The mean free path for consecutive spin flip events is called "spin diffusion length", and is much longer than that of electrons' collision[6]. Therefore, the diffusive transport of spins, i.e., the pure spin current, provides not only a variety of scientific interests but also an additional data transfer functionality for future spintronic device applications[7,8].

Hanle effect measurement is one of the most effective methods to characterize dynamic properties of the pure spin current[9,10]. When the magnetic field is applied perpendicular to the spin orientation, a collective spin precession is induced by its torque. In ballistic spin transport, spins can coherently rotate at a frequency proportional to the applied magnetic field. This allows us to control the direction of the spins in the channel and to manipulate the output signal of lateral spin valves (LSVs) by adjusting an effective external parameter such as the Rashba field tunable via a gate voltage[11]. This scheme realizes an active spin device such as the spin-transistor[12]. However, in a diffusive pure spin current in nonmagnets, the distributed transit times for individual spins cause the dephasing in the spin precession, and decreases drastically the spin accumulation in the Hanle effect measurment[13-22]. In this study, we employ dual spin injectors for LSVs to study coherency in the collective precession of the diffusive spins, which enhance the spin accumulation in the channel and also suppress the out of plane contribution to the Hanle effect signals due to the magnetization process of the ferromagnetic electrodes. Thereby we can detect a genuine in-plane precession signal in a 10 μm-long Ag wire which is much longer than its spin diffusion length. The coherent characteristics are investigated by the spin precession signal, revealing that the dwell time distribution narrows as the spins diffuse longer distance in the channel between the injector and the detector of LSVs.

**Results**

**Enhanced spin accumulation in lateral spin valves with dual injectors**

A conventional LSV consists of a pair of injector and detector ferromagnetic wires which are bridged by a nonmagnetic wire. The spins are injected by applying a bias voltage across the ferromagnetic/nonmagnetic interface and accumulate in its vicinity. Their density decays exponentially with a factor of $\exp(-d/\lambda_s)$ where $d$ is the distance from the injector and $\lambda_s$ is the spin diffusion length. Unlike the above LSV, our structure shown in Fig. 1a consists of three $Fe_{20}Ni_{80}$ (Permalloy) wires bridged by a Ag wire. The current $I$ is applied between FM1 and FM2 for the spin injection into the Ag wire, and the spin accumulation is detected in voltage $V$ by using FM3. To avoid the spin absorption by FM wires, we use the Permalloy/MgO/Ag junction in the present study[22,23]. For comparison between conventional and our schemes, we depict the spatial variation of spin accumulation $\delta\mu_{Ag}$ in the Ag wire calculated by using the same material parameters in Fig. 1b-d. In a conventional scheme of Fig. 1b, the spin injection generates the spin current $I_S$ towards both directions along the Ag wire, the magnitude of which is proportional to the spatial gradient of $\delta\mu_{Ag}$. On the other hand, our scheme shown in Fig. 1c and d, confines the spin current solely to the detector since the unnecessary side edge, i.e., relaxation volume, on the left of FM1 is removed. Figure 1c shows $\delta\mu_{Ag}$ from each injector with parallel magnetization configuration and the magnitude of each spin current from FM1 and FM2 can be twice as much as the conventional $I_S$ due to the confinement effect. As the direction of spin current across the FM1/MgO/Ag and FM2/MgO/Ag interfaces is opposite each other, the currents flowing across the interfaces thus cancel out for the parallel magnetic configuration of dual injectors, whereas for the anti-parallel configuration, the spin currents induced by FM1 and FM2 are constructive as depicted in Fig.1d. As a result, this scheme can enhance the total spin current by up to a factor of four compared to the conventional one.

The non-local spin valve signal in LSVs is shown in Fig. 2. For the conventional single injector LSV (SLSV), the high and low signals correspond respectively to the parallel and antiparallel configurations of the injector and detector FMs, of which overall change $\Delta R_S$ amounts to 31.5 m$\Omega$ at 10 K in Fig. 2a. Figure 2b shows the spin signal for the dual injector LSV (DLSV). All FMs exhibit distinct switching fields at around 10 mT, 40 mT and 50 mT, corresponding to the switching fields of FM1, FM2 and FM3, respectively. The hysteresis loop shows three-level signals associated with the following magnetic configurations; in the negative sweep (green line), the magnetization of FM1 flips at -10 mT and the injectors' magnetization configuration becomes antiparallel. Accordingly FM3 detects the change in the electrochemical potential and the resulting change in $\delta\mu_{Ag}$ is shown in Fig. 1d. Further decrease in field down to -40 mT flips the magnetization of FM2 and returns the injectors' magnetic configuration to parallel. Since the antiparallel configuration maximizes the spin accumulation in DLSV, it can be evaluated from the overall change $\Delta R_S$ between the minor loop shown in a red curve and the positive sweep shown in a blue line in Fig. 2b. The center-to-center injector and detector separation $L$ dependence of $\Delta R_S$ in Fig. 2c shows that $\Delta R_S$ of DLSV is almost 3 times larger than that of SLSV and decreases exponentially with increasing $L$ due to Elliot-Yafet spin relaxation mechanism in the Ag wire[24,25].

The analytical expression of $\Delta R_S$ for DLSV, for the case where the interface resistance is enough higher than the spin resistance of the nonmagnetic wire $R_N$, is approximated by using a solution of one-dimensional spin diffusion equation (see Supplementary Information for details)

$$\Delta R_S = \alpha P_I^2 R_N e^{-L/\lambda_N}, \tag{1}$$

where $\alpha = 1 + \exp(-2d_{12}/\lambda_N) + 2\exp(-d_{12}/\lambda_N)$ is the enhancement factor, $P_I$ is interfacial polarization, $\lambda_N$ is the spin diffusion length of the nonmagnet and $d_{12}$ is the separation between FM1 and FM2. The $\Delta R_S$ for DLSV is remarkably enhanced by a factor of $\alpha$ compared to that of SLSV, corresponding to the reduced equation (1) of $\Delta R_S$ for SLSV[26] with $d_{12} \gg \lambda_N$. The first and second terms in $\alpha$ represent the spin current injected from FM2 and the third term does the spin current injected from FM1. The obtained experimental results in Fig. 2c were fitted to equation (1) with adjusting parameters $P_I$ and $\lambda_N$, yielding $P_I = 0.36$, $\lambda_N = 1500$ nm and $\alpha = 3.2$. Note that the obtained valves of $P_I$ and $\lambda_N$ are consistent with our previous data for SLSV with Permalloy/MgO/Ag junctions[22] and $\alpha$ is decreased due to the spin relaxation in the channel between the two injectors.

**Experimental observation of genuine spin precession signal**

The Hanle effect measurements were performed on LSVs by applying perpendicular magnetic fields. Figure 3a shows the modulated non-local spin signal for SLSV and DLSV. A parabolic background signal is observed for the SLSV, the origin of which is the magnetization process of FMs. When the applied magnetic field is increased above the demagnetizing field of the FM wires, the magnetizations for the injector and the detector are tilted up along the field direction, pushing the background signal up towards the value of parallel configuration for FMs. To describe the both contributions of spin precession and magnetization process, we decompose them into that of spin precession in *x-y* plane and that of the *z* component reflecting the magnetization process. The non-local spin signal $V/I$ in the presence of $B_Z$ is thus given by the sum of the above two contributions;

$$\frac{V}{I} = R_S^{\text{Hanle}}\left(\omega_L, \mathbf{e}^{\text{FM1}}\cdot\mathbf{e}_y, \mathbf{e}^{\text{FM2}}\cdot\mathbf{e}_y, \mathbf{e}^{\text{FM3}}\cdot\mathbf{e}_y\right) + R_S^{\text{Hanle}}\left(0, \mathbf{e}^{\text{FM1}}\cdot\mathbf{e}_z, \mathbf{e}^{\text{FM2}}\cdot\mathbf{e}_z, \mathbf{e}^{\text{FM3}}\cdot\mathbf{e}_z\right), \tag{2}$$

with

$$R_S^{\text{Hanle}}(\omega_L, a^{\text{FM1}}, a^{\text{FM2}}, a^{\text{FM3}}) = \frac{1}{2}P_I^2 R_N \operatorname{Re}\left[\alpha_\omega\left(\lambda_\omega/\lambda_N\right)\exp(-L/\lambda_\omega)\right], \tag{3}$$

where $\alpha_\omega = a^{FM2}a^{FM3}\{1+\exp(-2d_{12}/\lambda_\omega)\} - 2a^{FM1}a^{FM3}\exp(-d_{12}/\lambda_\omega)$, $\lambda_\omega = \lambda_N/\sqrt{1+i\omega_L\tau_{sf}}$, $\tau_{sf} = \lambda_N^2/D_N$ the spin relaxation time, $D_N$ the diffusion constant, $\omega_L = \gamma_e B_z$ the Larmor frequency, $\gamma_e = g\mu_B/\hbar$ the gyromagnetic ratio, $g$ the g-factor, $\mu_B$ the Bohr magneton and $a^{FMi}$ is the projection of the unit vector of the magnetization of FMi $\mathbf{e}^{FMi}$ on y or z-axis. Note here that $R_S^{Hanle}(\omega_L, \mathbf{e}^{FM1}\cdot\mathbf{e}_y, \mathbf{e}^{FM2}\cdot\mathbf{e}_y, \mathbf{e}^{FM3}\cdot\mathbf{e}_y)$ represents the nonlocal resistance at the precessional frequency $\omega_L$, and $R_S^{Hanle}(0, \mathbf{e}^{FM1}\cdot\mathbf{e}_z, \mathbf{e}^{FM2}\cdot\mathbf{e}_z, \mathbf{e}^{FM3}\cdot\mathbf{e}_z)$ does that without spin precession (see Supplementary Information for details). For SLSV, we obtain $P_I = 0.37$ and $\lambda_N = 1420$ nm by fitting equation (2) to the experimental data as shown in Fig. 3a, which are consistent with those obtained alternatively from the $L$ dependence of $\Delta R_S$ in the previous section. For DLSV, the z component of the injectors is canceled out because of the opposite direction of the applied current to the junctions as depicted in Fig. 3b. This allows us to detect the genuine precession signal in Fig. 3a.

**Towards coherent spin precession**

In the diffusive pure-spin transport, the collective spin precession decoheres due to broadening of the dwell time distribution in the channel between the injector and the detector[10]. The amplitude of the spin valve signal at $B_Z = 0$ decreases after the $\pi$ rotation at $B_Z^\pi = 0.16$ T, as can be seen in Fig. 3a. In order to better quantify the coherency in the collective spin precession, we define the figure of merit as the ratio $\Delta R_S^\pi / \Delta R_S^0$, where $\Delta R_S^\pi$ and $\Delta R_S^0$ are respectively the amplitude of the spin valve signal right after the π rotation and that in zero field right before the rotation begins. The $\Delta R_S^\pi / \Delta R_S^0$ increases with increasing $L$, and the experimental trend is well reproduced by equation (2) as shown in Fig. 3c. To understand the observed trend in more detail, we employ the one-dimensional diffusion model which gives the y-component of net spin density at the detector $\langle S_y \rangle \propto 1/\sqrt{4\pi D_N t} \exp\left(-\left(L^2/4D_N t\right) - t/\tau_{sf}\right)\cos(\omega_L t)$, as a function of the dwell time $t$ in the presence of $B_Z$[10]. The $\langle S_y \rangle$ versus $t$ curves for $L = \lambda_N$ with $B_Z = 0$ and $B_Z = B_Z^\pi$ are shown in Fig. 3d. When $B_Z = 0$, $\langle S_y \rangle$ takes a broad peak structure followed by a long exponential tail. The detected spin signal in LSVs is proportional to the $\langle S_y \rangle$ integrated over time. The distribution of $\langle S_y \rangle|_{B_z=0}$ gets narrower as the channel length becomes longer, of which evolution is depicted in three distribution curves under $B_Z = 0$ of Figs. 3d-f. The long exponential tail observed in Fig. 3d diminishes in proportion to $1/\sqrt{t}\exp(-t/\tau_{sf})$. When $B_Z = B_Z^\pi$ is applied, the integrated value of $\langle S_y \rangle$ over time cancel for short separation $L \sim \lambda_N$ (Fig. 3d) whereas it does not cancel for long separation $L \gg \lambda_N$ (Figs. 3e and f), indicating that the coherence of collective spin precession is well preserved for long spin transport. This trend is experimentally observed as an increase of $\Delta R_S^\pi / \Delta R_S^0$ from 0.21 to 0.53 with $L$ as shown in Fig. 3c. In spin-polarized electron transport, a drift field is utilized to suppress the distributed transit time of the spins and almost coherent spin precession with $\Delta R_S^\pi / \Delta R_S^0 \sim 1$ is reported for a 350 μm-thick undoped single-crystal Si wafer [27].

**Discussion**

To better understand the coherence in collective spin precession for the pure-spin current, $t = \tau_{sf}T$ is substituted into the distribution function at $B_Z = 0$. We then obtain $<S_y> \propto 1/\sqrt{T} \exp\left(-(L/2\lambda_N)^2/T - T\right)$, where $T$ is dimensionless time. This implies that the distribution of the dwell time, i.e., coherency, is characterized only by $L/\lambda_N$ and more importantly it does not depend on the kind of materials as long as their transport is diffusive. To check this idea the $L/\lambda_N$ dependence of $\Delta R_S^\pi/\Delta R_S^0$ are summarized by using the data so far reported for metals, semiconductors and graphene in Fig. 4. Interestingly the relation between the coherence and the normalized separation shows a universal behavior and the experimental data are well reproduced by equation (2). We shall note here that the effective length $L/\lambda_N$, not the spin lifetime, is an important parameter to manipulate the spin precession coherently in the diffusive pure-spin transport while the spin accumulation is relaxed during the diffusive transport in the channel. Therefore, the high spin injection efficiency of the Permalloy/MgO/Ag junction and the confinement effect in the DLSV structure could offer advantages for realizing giant spin accumulation as well as coherent spin precession along a 10 μm-long Ag wire which is much longer than the spin diffusion length.

For spintronic devices using such a long-diffusion spin current, fast spin transport may be critical. The high coherence of the spin precession over $\Delta R_S^\pi/\Delta R_S^0 = 0.4$ is reported for Al and graphene as can be seen in Fig. 4, however, the diffusion constant of the pure spin current is 0.003 m$^2$/s and 0.01 m$^2$/s [13,20], respectively, which is much slower than that of Ag (0.047 m$^2$/s). Therefore, the experimental results in this study could be useful in developing a new class of spintronic devices and the material-independent perspective for the spin precession will be beneficial for us to design pure-spin-current-based memory and logic devices by using a variety of metallic and semi-conductive materials including graphene.

**Methods**

Lateral spin valves with Permalloy/MgO/Ag junctions are prepared on a Si/SiO$_2$ substrate by means of shadow evaporation using a suspended resist mask which is patterned by e-beam lithography. All the layers are e-beam deposited in an ultra-high vacuum condition of about 10$^{-6}$ Pa. First, 20-nm-thick Permalloy layer is obliquely deposited at a tilting angle of 45° from substrate normal. Second, the interface MgO layer is deposited at the same tilting angle of 45°. Third, 100-nm-thick Ag layer is obliquely deposited normal to the Si substrate. Finally, 3-nm-thick capping MgO layer is deposited to prevent surface contamination of the devices. After the liftoff process, the devices are annealed at 400 °C for 40 min in an N$_2$ (97%) + H$_2$ (3%) atmosphere [28].

The non-local measurements are carried out by a dc current source and nano-voltmeter. The bias current in the range between 200 and 400 µA is applied to the injector. The magnetic field is applied parallel to the Permalloy wires for the spin valve measurements. The switching field of Permalloy wires is controlled not only by changing the width but also by attaching a large domain wall reservoir at the edge. For the Hanle effect measurements, the magnetic field is applied perpendicular to the Si substrate. The field direction is carefully controlled to rule out the misalignment that causes the in-plane field component that switches the magnetization of Permalloy in the plane during the measurements.

**Acknowledgements**

This work is partly supported by Grant-in-Aid for Scientific Research (A) (No. 23244071) and Young Science Aid (A) (No. 23681032) from the Ministry of Education, Culture, Sports, Science and Technology, Japan.

**Additional information**

The authors declare no competing financial interests.

**Figure Legends**

Figure 1. **Sample structure and spatial variation of spin accumulation. a**, Schematic diagram of lateral spin valve with dual injectors in non-local measurement configuration. The Permalloy wires are 140 nm in width and 20 nm in thickness. The Ag wire is 120 nm in width and 100 nm in thickness. Center-to-center separation between FM1 and FM2 $d_{12}$ is 350 nm. **b,** Schematic diagram of LSV with single injector and the spatial variation of spin accumulation $\delta\mu$ for Ag. Arrows in Ag and FM respectively represent the non-equilibrium magnetization of Ag and the magnetization of FM. The spin current $I_S = I_\uparrow - I_\downarrow$ flows in both directions along the Ag wire. The magnitude of $I_S$ is proportional to the spatial gradient of $\delta\mu_{Ag}$. **c,d,** Schematic diagram of LSV with dual injectors with parallel or anti-parallel configuration and the spatial variation of spin accumulation $\delta\mu$ for Ag. The red and blue lines respectively show $\delta\mu_{Ag}$ induced by spin injectors of FM1 and FM2. In parallel configuration, the flow direction of spin current from FM2 $I_{S2}$ is opposite to that of FM1 $I_{S1}$. In anti-parallel configuration, the flow direction of $I_{S1}$ is the same as that of $I_{S2}$.

Figure 2. **Non-local spin signals. a**, Spin signal as a function of y-directional magnetic field for SLSV at 10 K. The arrows indicate the magnetization configuration of two Permalloy wires. The triangles indicate the sweep direction of the magnetic field. **b**, Spin signal as a function of magnetic field for DLSV at 10 K. The arrows indicate the magnetization configuration of three Permalloy wires. The triangles indicate the sweep direction of the magnetic field. The red line shows the minor hysteresis loop. **c**, Spin valve signal $\Delta R_S$ as a function of injector-detector separation at T = 10 K. The solid lines are the fitting curves using equation (1) with the same fitting parameters for SLSV and DLSV.

Figure 3. **Spin precession measurements by using Hanle effect. a**, Non-local spin signal modulated by spin precession as a function of perpendicular field for single and dual injector LSVs with $L$ = 6 μm at 10 K. The solid lines are the fitting curves using equation (2). **b**, Schematic diagram of magnetizations of injectors and non-equilibrium magnetization *m* in the Ag wire in the presence of high $B_Z$. In Ag, *y*-components of *m*1 and *m*2 are constructive but *z*-component of them are canceled each other out both for anti-parallel and parallel magnetization configurations of the injectors. **c**, Coherent parameter of the spin precession $\Delta R_S^\pi / \Delta R_S^0$ as a function of $L$. The solid lines are the fitting curves using equation (2) with the same parameters as used in **a**. Data are corrected by taking account of the influence of magnetization process. **d-f**, Density of *y*-directional spin arrived at the detector as a function of dwell time for LSVs with different $L$. The black and red lines, respectively, represent the distribution of the dwell time in the channel without and with $B_Z$ that causes π rotation after spins are injected.

Figure 4. **Coherency of spin precession in diffusive pure-spin current.** Coherent parameter of the spin precession $\Delta R_S^{\pi}/\Delta R_S^{0}$ as a function of $L/\lambda_\mathrm{N}$. The solid line is a universal curve obtained from equation (2).

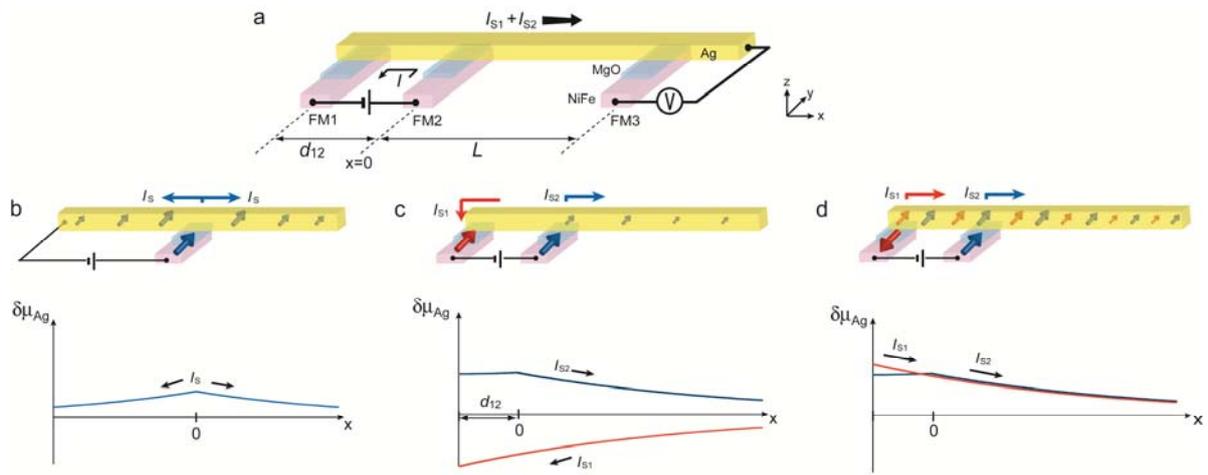

Figure 1, Idzuchi *et al*

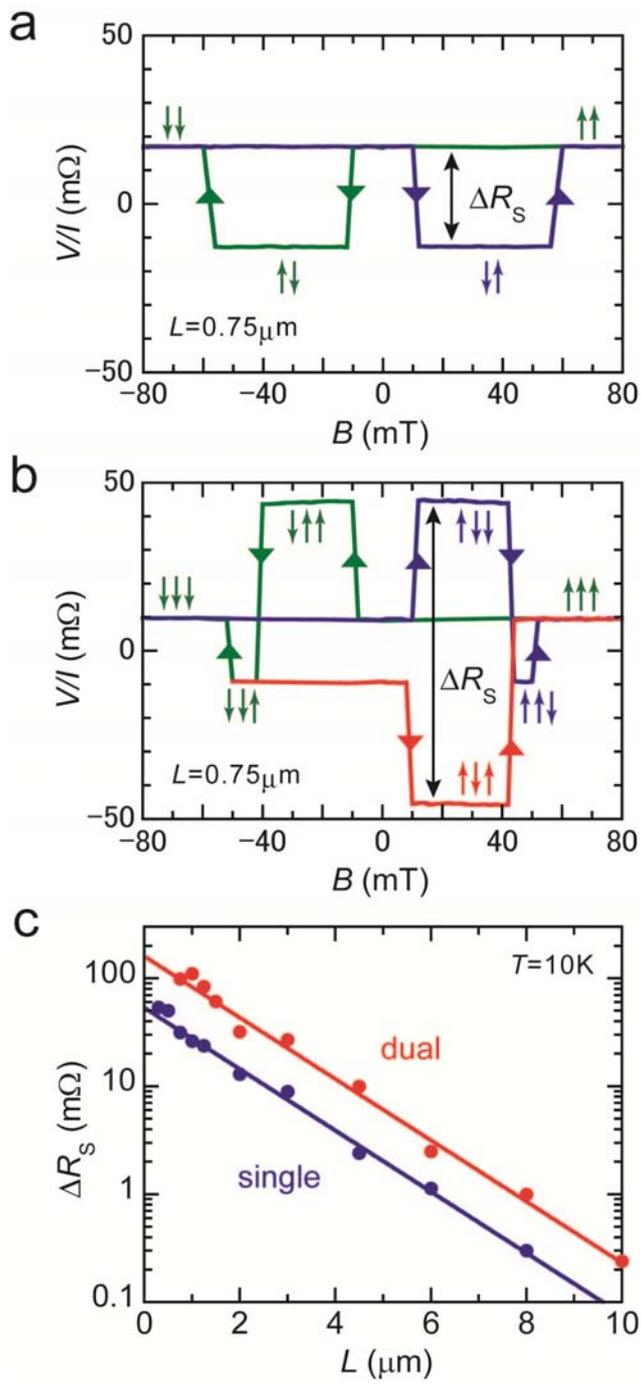

Figure 2, Idzuchi *et al*

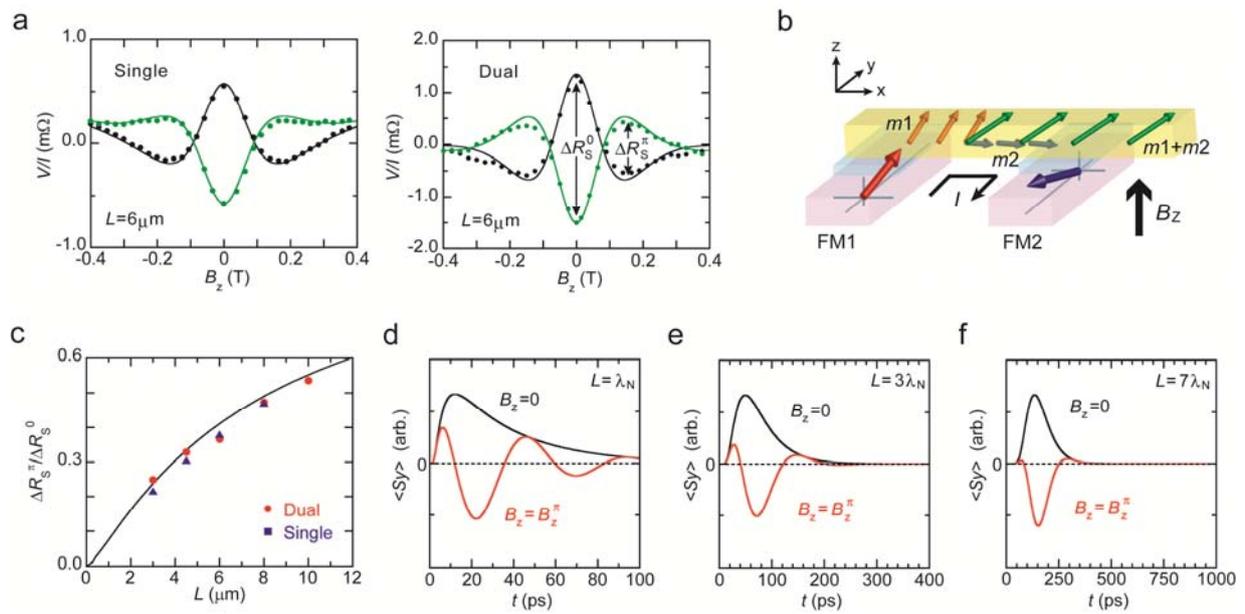

Figure 3, Idzuchi *et al*

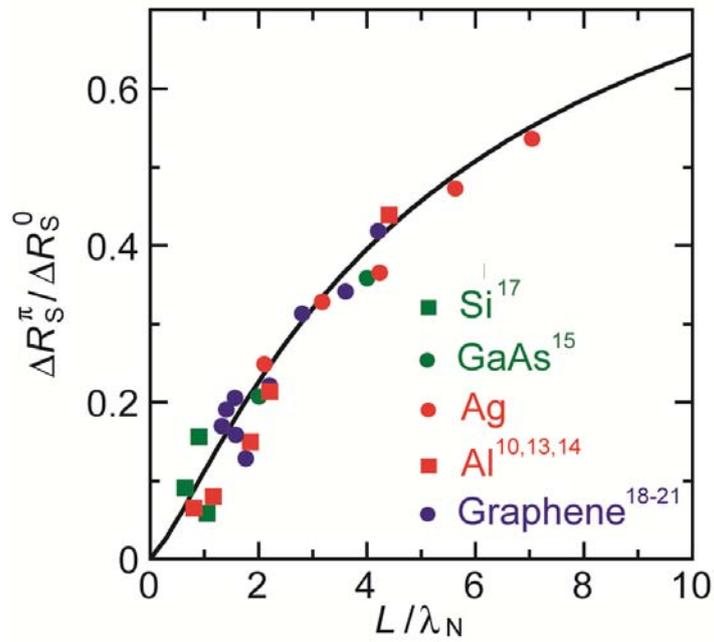

Figure 4, Idzuchi *et al*